\documentstyle[preprint,aps]{revtex}

\begin{document}

%\draft

\preprint{DUKE-TH-95-90}

\title{Chaos Driven by Soft-Hard Mode Coupling in Thermal Yang-Mills Theory}

\author{T. S. Bir\'o\footnote{Theory Division, MTA KFKI, RMKI,
Pf. 49, H-1525 Budapest 114, Hungary},
 S. G. Matinyan\footnote{On leave of absence from Yerevan Institute of
Physics, Armenia},
 and B. M\"uller}

\address{Department of Physics, Duke University, Box 90305,
Durham, NC 27708-0305}

\date{Version \today}
\maketitle

\begin{abstract}
We argue on a basis of a simple few mode model of SU(2) Yang-Mills theory
that the color off-diagonal coupling of the soft plasmon to hard thermal
excitations of the gauge field drives the collective plasma oscillations
into chaotic motion despite the presence of the plasmon mass.
\end{abstract}


The chaoticity of classical nonabelian gauge fields was originally
discovered for spatially homogeneous gauge potentials \cite{r1} and
later extended to special, radially symmetric field configurations
\cite{r1a,r1b}.  More recently, the chaotic dynamics of the full
classical Yang-Mills equations was investigated in numerical studies
of Hamiltonian SU($N_c$) lattice gauge theory with $N_c=2,3$ color
degrees of freedom \cite{r2,r2a}.  The complete Lyapunov spectrum of
the SU(2) gauge field was calculated \cite{r3a}, and it was found
that the field configurations ergodically cover the available phase
space on the time scale of the inverse of the leading Lyapunov
exponent, $\tau \approx 1/\lambda_0 \propto 1/g^2E$, where $E$ is the
energy per lattice degree of freedom.  The leading Lyapunov exponent
coincides with twice the plasmon damping rate obtained in resummed
perturbation theory at finite temperature $T$ \cite{r4}.

The coupling to the heat bath gives rise to a mean
inertia of the plasmon field. For long wavelength ($k\to 0$) and static
fields $(\omega =0$) it coincides with the Debye mass square $m^2_{\rm P}
= \frac{N_c}{3} g^2T^2$;  for on-shell plasmons in the limit $k \to 0$
one has $m_{\rm P}^2 = \frac{N_c}{9} g^2T^2$.  Such a mass restricts the
region of chaotic motion to large amplitude plasmon fields. In fact a
control parameter,
\begin{equation}
R = {m_{\rm P}^4 V \over 16g^2E}
\end{equation}
with $E$ being the total classical energy of the plasmon field
and $V$ the plasma volume, has been found to control chaos in
simple, few-mode models of the Yang-Mills field theory\cite{r5}.
For small $R \le 0.2$ chaos develops.  If the mean plasmon mass
$m_{\rm P}$ were the sole effect of hard thermal gluons, then small
amplitude plasmon fields would not grow chaotic \cite{r7}.

We now argue that even small amplitude collective plasma oscillations
can be chaotic because of the color-nondiagonal coupling to the
fluctuating components of the plasmon self-energy.  These color
non-diagonal matrix elements can render some eigenvalues of the mass
square matrix $M_{\rm P}$ negative.  This is a source of instability for
small amplitude soft fields.  Once the energy stored in the soft
components has grown, the classical chaotic behavior sets in.

We are not able yet to prove the above scenario generally.  In this
paper we study as an example the coupled dynamics of some selected
Fourier components of the vector potential to show how the mechanism
discussed above can be realized.  The central point here is the
existence of a hierarchy of relevant timescales in the dynamics of
oscillations of the gauge field.  In fact, three such timescales can
be identified.\footnote{We here assume, as is customary in the analysis
of thermal gauge theories, that $g$ is sufficiently small that the
hierarchy $t_1\ll t_2\ll t_3$ holds.}  The hard thermal modes oscillate
over a period of $t_1 \sim 1/T$.  Due to the nonabelian coupling they
influence the soft mode dynamics on the timescale $t_2 \sim 1/gT$.
This also sets the timescale for the evolution of the soft plasma
oscillations.  Finally, the thermal ensemble average over particular
hard mode configurations is characterized by the timescale of the gluon
damping rate $t_3 \sim 1/g^2T$.  The fact that $t_2 \le t_3$ requires
that the ensemble average has to be taken over solutions of the plasmon
equation of motion rather than over the equation itself.

Let us now consider a simplified model retaining the essential features.
Here we build on the infrared limit of Yang-Mills field theory
(``Yang-Mills mechanics''\cite{r1}).  Each Fourier mode of the SU(2)
gauge field representing hard thermal gluons carries six degrees of
freedom (two polarizations, three internal degrees of freedom).  Soft
modes, neglecting spatial derivatives, can be described by the three
eigenvalues of the $O(3) \times O(3)$---symmetric vector potential
$A_i^a$.  Without loss of generality they can be chosen as \cite{Sav}:
\begin{equation}
x = A^1_1, \qquad y = A^2_2, \qquad z = A^3_3.
\end{equation}
The most general ansatz describing one hard and one soft momentum contains
therefore nine coupled modes
\begin{eqnarray}
A^1_1 = x + w^1_1, \qquad & A^1_2 = w^1_2, & \qquad A^1_3 = 0,
\nonumber \\
A^2_1 = w^2_1, \qquad & A^2_2 = y + w^2_2, & \qquad A^2_3 = 0,
\nonumber \\
A^3_1 = w^3_1, \qquad & A^3_2 = w^3_2, & \qquad A^3_3 = z.
\end{eqnarray}
Here
\begin{equation}
w_i^a = \sqrt{2} \left( u_i^a \cos (\vec{k}\cdot\vec{r})
  + v_i^a \sin (\vec{k}\cdot\vec{r}) \right)
\end{equation}
denote the hard thermal modes with time-dependent amplitudes $u_i^a$
and $v_i^a$.  We choose the thermal wave number $\vec{k}=T\vec e_3$ to
point in the third direction.  Then all longitudinal components $w^a_3=0$
vanish and only one derivative (in the third direction) differs from zero
in this ansatz:
\begin{equation}
\partial_3 w_i^a = T \sqrt{2} \left( - u_i^a
   \sin (\vec{k}\cdot\vec{r})
   + v_i^a \cos (\vec{k}\cdot\vec{r}) \right).
\end{equation}
Choosing the temporal gauge, $A_0^a=0$, the electric field components are
simply the time derivatives of the vector potential
\begin{eqnarray}
E^1_1 = \dot{x} + \dot{w}^1_1,
\qquad & E^1_2 = \dot{w}^1_2, & \qquad E^1_3 = 0,
\nonumber \\
E^2_1 = \dot{w}^2_1,
\qquad & E^2_2 = \dot{y} + \dot{w}^2_2, & \qquad E^2_3 = 0,
\nonumber \\
E^3_1 = \dot{w}^3_1, \qquad & E^3_2 = \dot{w}^3_2, & \qquad E^3_3 = \dot z.
\end{eqnarray}
The nonabelian magnetic field has the following components
\begin{eqnarray}
B^1_1 &=& -\partial_3 w^1_2 + gz(y+w^2_2),
\nonumber \\
B^1_2 &=& \partial_3 w^1_1 - gzw^2_1,
\nonumber \\
B^1_3 &=& g(w^2_1w^3_2-w^2_2w^3_1) - gyw^3_1,
\nonumber \\
B^2_1 &=& -\partial_3 w^2_2 - gzw^1_2,
\nonumber \\
B^2_2 &=& \partial_3 w^2_1 + gz(x+w^1_1),
\nonumber \\
B^2_3 &=& g(w^3_1w^1_2-w^3_2w^1_1) - gxw^3_2,
\nonumber \\
B^3_1 &=& -\partial_3 w^3_2,
\nonumber \\
B^3_2 &=& \partial_3 w^3_1,
\nonumber \\
B^3_3 &=& g(w^1_1w^2_2-w^1_2w^2_1) + g(w^1_1y+w^2_2x) + gxy.
\end{eqnarray}
In order to obtain the effective Hamiltonian for the soft modes, which are
taken to be constant throughout space, we perform a spatial average over
the relevant terms in the Hamiltonian.  For the quadratic and quartic
expressions of the hard amplitudes $w_i^a$ we use the following relations
\begin{eqnarray}
&&\overline{w_i^aw_j^b} = u_i^au_j^b + v_i^av_j^b,
\nonumber \\
&&\overline{w_i^a\partial_3 w_j^b} =
  T \left( u_i^av_j^b - v_i^au_j^b \right),
\nonumber \\
&&\overline{\partial_3 w_i^a \partial_3 w_j^b} =
  T^2 \left( u_i^au_j^b + v_i^av_j^b \right),
\nonumber \\
&&\overline{w_i^aw_j^bw_i^aw_j^b} =
{3 \over 2} \left( u_i^au_i^au_j^bu_j^b + v_i^av_i^av_j^bv_j^b
+ u_i^av_i^au_j^bv_j^b + u_i^au_i^av_j^bv_j^b \right).
\end{eqnarray}
The spatially averaged Hamiltonian,
\begin{equation}
\overline{H} = {1 \over 2} \overline{E_i^aE_i^a} +
{1 \over 2} \overline{B_i^aB_i^a} \equiv H_T+H_S+H_I
\end{equation}
contains three parts:  $H_T$, describing the hard thermal modes
interacting among themselves;  $H_S$, describing the soft modes in
isolation; and $H_I$ denoting the interactions between hard and soft
modes.  According to the usual picture of a gluon plasma we treat the
hard modes perturbatively, neglecting their mutual interactions.  This
yields:
\begin{equation}
H_T \longrightarrow H_0 = {1 \over 2} \sum_{i=1}^2 \sum_{a=1}^3
\left[\left( \dot{u}_i^a\dot{u}_i^a + \dot{v}_i^a\dot{v}_i^a \right)
+ T^2 \left( u_i^au_i^a + v_i^av_i^a \right)\right],
\label{Ham}
\end{equation}
in the leading order approximation.

The part of the Hamiltonian which contains soft modes must be treated
non-perturbatively, so we keep the quartic anharmonic terms here. We get
\begin{equation}
H_S = {1 \over 2} \left( \dot{x}^2 + \dot{y}^2 + \dot{z}^2 \right)
 + {g^2 \over 2} \left( x^2y^2 + y^2z^2 + z^2x^2 \right)
\label{AH}
\end{equation}
for the soft mode part and
\begin{equation}
H_I = {1 \over 2} \left( M_1^1 x^2 + M^2_2 y^2 + (M_1^2+M_2^1)xy
+ M^3_3 z^2 \right) + Q^3z
\end{equation}
for the interaction between soft and hard thermal modes.  Here
\begin{eqnarray}
&&M_1^1 = g^2 \left( u^3_2u^3_2 + v^3_2v^3_2
   + u^2_2u^2_2 + v^2_2v^2_2  \right),
\nonumber \\
&&M_2^2 = g^2 \left( u^3_1u^3_1 + v^3_1v^3_1
   + u^1_1u^1_1 + v^1_1v^1_1  \right),
\nonumber \\
&&M_1^2 = M_2^1 = g^2 \left( 2u^1_1u^2_2 + 2v^1_1v^2_2
   - u^1_2u^2_1  - v^1_2v^2_1 \right),
\nonumber \\
&&M_3^3 = g^2 \left( u^1_1u^1_1+v^1_1v^1_1+u^1_2u^1_2+v^1_2v^1_2
       +u^2_1u^2_1+v^2_1v^2_1+u^2_2u^2_2+v^2_2v^2_2 \right),
\nonumber \\
&&Q^3 = 2gT\left( u^1_1v^2_1-u^2_2v^1_2+u^1_2v^2_2-u^2_1v^1_1 \right)
\label{MHam}
\end{eqnarray}
are the components of the effective mass square matrix for the soft
modes and $Q^3$ is the fluctuating color charge of the hard modes,
which acts as a source for the soft modes.

The Hamiltonian $H_0$ (\ref{Ham}) describes independent oscillations with
a frequency of $\omega = |\vec{k}| = T$. For the dynamical evolution
of the soft modes $x, y, z$ this is a very fast motion; we can
therefore average over many time periods in order to arrive at an
effective soft mode Hamiltonian $\overline{H_I}$ for the interaction.  The
ensemble averaging over the relative phases and slowly fluctuating
amplitudes of the hard thermal components, however, cannot be done at
this point, because these quantities change on the timescale $t_3\sim
1/g^2T$.  Consider the hard thermal eigenmodes of the Hamiltonian
(\ref{Ham}):
\begin{eqnarray}
&& u_i^a(t) = U_i^a \cos (Tt+\varphi^a_i),
\nonumber \\
&& v_i^a(t) = V_i^a \cos (Tt+\psi^a_i).
\label{EH}
\end{eqnarray}
The initial amplitudes $U_i^a, V_i^a$ and phases $\varphi^a_i$,
$\psi^a_i$ ($i=1,2$ $a=1,2,3$) determine the quantities $\bar{M_i^a}$ and
$\bar{Q^a}$ which in turn influence the evolution of the soft modes $x,y,z$.
In order to obtain the correct scaling $\bar{M_i^a}\sim g^2T^2$ of the
elements of the mass matrix (\ref{MHam}) we take the amplitudes of the fast
oscillations to be ${\cal O}(T)$.  As seen from (\ref{Ham}) this is
consistent with an energy density $\bar H_0\sim T^4$ in the hard modes,
as expected for the thermal bath.

If, for the sake of simplicity, we assume all amplitudes equal to
each other, so $U_i^a=V_i^a\equiv U\sim T$, we obtain $\bar{M_1^1} =
\bar{M_2^2} = m^2$, $\bar{M_3^3} = 2m^2$ with
\begin{equation}
m^2 = 2g^2U^2 \sim g^2T^2.
\end{equation}
These effective mass terms in the averaged Hamiltonian $\bar H_I$ are
the analogue of the plasmon mass term in the full thermal gauge theory.
For small amplitudes of the soft modes $x,y,z$ they dominate over the
anharmonic terms in (\ref{AH}) and apparently cause the soft modes to
oscillate harmonically with a frequency $\omega_S\sim gT$.

However, the presence of non-diagonal elements $M_1^2$ in the mass square
matrix can destabilize these oscillations.  The time averaged off-diagonal
term $\overline{M_1^2}$, as well as the source term $\overline{Q^3}$,
depend on the phases of the hard thermal oscillations (\ref{EH}):
\begin{equation}
\overline{M_1^2} = {m^2 \over 2} \left( \cos(\varphi^1_1 - \varphi^2_2) +
\cos(\psi^1_1 - \psi^2_2) - {1 \over 2}\cos(\varphi^1_2 -
\varphi_1^2) - {1 \over 2}\cos(\psi^1_2 - \psi^2_1) \right),
\end{equation}
and
\begin{equation}
\overline{Q^3} = {{m^2T} \over {2g}} \left(  \cos(\varphi^1_1-\psi^2_1)
-\cos(\varphi^2_2-\psi^1_2)
+\cos(\varphi^1_2-\psi^2_2)
-\cos(\varphi^2_1-\psi^1_1) \right).
\end{equation}

Because of their dependence on the relative phases of the fast modes
$\overline{M_1^2}$ and $\overline{Q^3}$ would vanish if a complete
ensemble average over the fast modes were performed.  However, as pointed
out at the beginning of our discussion, the phases and amplitudes of
these modes change more slowly (on timescale $t_3$) than the soft modes
oscillate (on timescale $t_2$).  Hence, both $\overline{M_1^2}$ and
$\overline{Q^3}$ must be
considered as secular quantities in the averaged (over timescale $t_1$)
Hamiltonian $H_S+\overline{H_I}$ governing the evolution of the soft modes.

Far from the ``average'' case, $\overline{M_1^2} = \overline{Q^3} = 0$,
there is the extreme possibility of $\overline{M^1_2} = 3m^2/2$ and
$\overline{Q^3} =m^2T/2g \sim gT^3$ obtained for
$\varphi_1^1=\varphi_2^2=\psi_1^2
=\psi_2^1\pm\pi$, $\psi_1^1=\psi_2^2=\varphi_2^1= \varphi_1^2\pm\pi$.
As we will see below, in this case the mass square matrix of the coupled
$x$ and $y$ modes has negative eigenvalues giving rise to an {\em
exponentially growing} solution of the linearized equations of motion for
small amplitudes $x$ and $y$.

We now consider small amplitude oscillations and neglect terms nonlinear
in $x$ or $y$.  The $z$-coordinate then decouples from the $x-y$ motion,
as seen from (\ref{MHam}).  It is most convenient to cast the resulting
equation into matrix form
\begin{equation}
{{d^2\over dt^2}+ \overline{M_1^1} \qquad\quad \overline{M_1^2} \choose
\overline{M_1^2} \qquad\quad {d^2\over dt^2} + \overline{M_2^2}}
{x\choose y} = {0\choose 0},  \label{MatH}
\end{equation}
where we remind the reader that the ``bar'' denotes the time averaged
quantities for fixed phase $\varphi_i^a,\psi_i^a$ and amplitudes $U_i^a,
V_i^a$.  Since the average over the ensemble of these initial conditions
must be performed after the full solution is obtained, we proceed with
solving (\ref{MatH})  by finding the eigenvalues and eigenvectors of the
matrix operator on the left hand side.

Seeking the solution with $x,y\propto e^{i\omega t}$, we find the
eigenvalue equation
\begin{equation}
\omega^4 -\omega^2 \left( \overline{M_1^1}+\overline{M_2^2}\right) +
\left( \overline{M_1^1}\, \overline{M_2^2} - \overline{M_1^2}\,
\overline{M_2^1}\right) = 0.
\label{barM}
\end{equation}
Using the previous result $\overline{M_1^1} = \overline{M_2^2} = m^2$ and
$\overline{M_1^2}=\overline{M_2^1} = \gamma m^2$ with $\vert\gamma\vert
\le 3/2$, we find
\begin{equation}
\omega_{\pm}^2 = m^2(1\pm\vert\gamma\vert). \label{vertm}
\end{equation}
Clearly, for $\vert\gamma\vert >1$ one of the eigenvalues becomes unstable
with an imaginary frequency.

With the help of the eigenvalues $\omega_{\pm}$ and eigenvectors
$(x_{\pm},y_{\pm})$ the general solution of the time-averaged equation
(\ref{MHam}) can be written in the form
\begin{equation}
{x\choose y} = \Delta (t) {\dot x (0)\choose \dot y (0)} + \dot\Delta(t)
{x(0)\choose y(0)},
\end{equation}
where
\begin{equation}
\Delta (t) = \sum_{i=\pm} {\sin\omega_it\over \omega_i} {c_i^2\quad c_id_i
\choose c_id_i \quad d_i^2}.  \label{22}
\end{equation}
where $(c_i,d_i)$ are the eigenvectors corresponding to the eigenvalues
$\omega_i^2$.  It is not difficult to show that the product $c_id_i$
vanishes when it is averaged over the phases $\varphi_i^a$ and $\psi_i^a$.
As a result, the full thermal ensemble average gives rise to a color
diagonal effective soft mode propagator $\langle\Delta (t)\rangle$,
which contains an exponentially growing part originating from those
regions of phase space where $\omega_-^2 <0$, as is seen from (\ref{22}).

Summarizing, we demonstrated in a simple model that the
coupling of soft plasmon oscillations to hard thermal gluons can drive
the soft field amplitude into an exponential growth and with that into
a chaotic dynamical behavior.  There are, of course, many possible few
mode approximations to the Yang-Mills field theory.  The example we
discussed above was generic in the sense that it showed both dynamical
mass generation and  the development of chaotic instability due to
off-diagonal parts of the self-energy, as well as source term coupling
to the soft modes.  One may expect that by this mechanism the dynamics
of long wavelength modes in a hot gluon plasma remains chaotic despite
the presence of a dynamical plasmon mass.

\acknowledgements

This work was supported in part by the U.S. Department of Energy
(grant {\sc de-fg05-90er40592}) and in part by Hungarian
National Scientific Research Fund OTKA (grant {\sc T-014213}).
S.G.M. thanks Prof. V. Gregorian, President of Brown University,
and the Office of International Programs of Duke University
for their support.

\end{document}